\documentclass[twoside]{article}

\usepackage{lipsum} 
\usepackage{aas_macros}

\usepackage[sc]{mathpazo} 
\usepackage[T1]{fontenc} 
\usepackage{microtype} 

\usepackage[hmarginratio=1:1,top=32mm,columnsep=20pt]{geometry} 
\usepackage{multicol} 
\usepackage[hang, small,labelfont=bf,up,textfont=it,up]{caption} 
\usepackage{booktabs} 
\usepackage{float} 
\usepackage{hyperref} 
\usepackage{natbib} 
\usepackage{graphicx}
\usepackage{caption} 
\usepackage{subcaption} 
\usepackage{lettrine} 
\usepackage{paralist} 
\usepackage{wasysym}
\usepackage{stmaryrd}

\usepackage{abstract} 

\usepackage{color}

\usepackage{ulem}

\usepackage{titlesec} 
\renewcommand\thesection{\Roman{section}} 
\renewcommand\thesubsection{\thesection.\arabic{subsection}} 
\titleformat{\section}[block]{\large\scshape\centering}{\thesection.}{1em}{} 
\titleformat{\subsection}[block]{\large}{\thesubsection.}{1em}{} 

\usepackage{fancyhdr} 

\usepackage{xurl} 
\usepackage{color}

\title{\vspace{-15mm}\fontsize{24pt}{10pt}\selectfont\textbf{Astronomers Getting Less Creative Over Time Is Why This Title Isn't Better}} 

\author{
\large
\textsc{Michael B. Lund$^1$}\\
\normalsize $^1$Caltech/IPAC-NExScI\\
\normalsize \href{mailto:editor@actaprimaaprilia.com}{editor@actaprimaaprilia.com} 
\vspace{-5mm}
}
\date{}

\begin{document}

\maketitle 

\thispagestyle{fancy} 


\begin{abstract}
For generations, people have complained that things used to be better in the past. In this paper, we investigate this change by specifically looking at creativity in astronomy. To do this,we explore if older constellations reflected a greater sense of creativity on the part of those designing them than more modern constellations do. We find that things really have become simplistic and less original over time.
\end{abstract}


\begin{multicols}{2} 

\section{Introduction}
\lettrine[nindent=0em,lines=3]{F}or the ancient Greeks, who helped shape the foundation of modern western thought, astronomy had the now-somewhat-surprising position of having its own Muse, Urania. While a modern reader would more likely associate the Muses with the arts, \citet{Knishkowy2024} makes the case that to the Greeks, there was little difference between artistic and scientific inspiration:
\begin{quote}
    The role of the Muses in Ancient Greek religion and culture illustrates the civilization’s
view that accomplishment in any valued intellectual field, whether artistic or otherwise, fell under
the nine deities’ domain of divine inspiration. This implies that, to the ancient Greeks, there was
no such thing as an uncreative intellectual pursuit, and that the line between art and science was
blurred, if it existed at all...it is evident that the ancient Greeks viewed creativity and reason as
two vital elements in the same processes that led to great works in theater, poetry, astronomy, and
beyond.
\end{quote}

Since then, the view that things are less creative, imaginative, and unoriginal has run rampant. Photography has been derided as inferior to paintings and engravings and too mechanical to have creativity of its own present \citep{Teicher2016}. Studies of music have shown that contemporary Western music has been homoginizing over the last few decades, becoming increasingly similar \citep{Serra2012}. Critics of contemporary film have said that "[o]riginality is at an all-time low" \citep{Shirazi2024}. Most broadly, tests that are designed to measure people's creativity have also shown a decline in the population's creativity over time \citep{Kim2011}.

Astronomy has not been spared of this criticism, as many outside astronomical research have observed the pattern of uncreative nomenclature that begins with terms like "red giant" for stars that are big and red and "white dwarf" for stars that are white and small. In 2001, \citeauthor{CBCNews2001} covered that the IAU was making it clear that stars would only have 'boring names' rather than the creative names that we have inherited from sources like Greek and Arabic culture. The same pattern was noticed extending to planets in 2015 by \citeauthor{Stromberg2015}, complaining that despite so many fascinating new planets being discovered, "they all have utterly dull, unmemorable names". Astronomers have even been accused of not being able to name nothing right \citep{Keim2007}.
But perhaps no single indictment is stronger than the news coverage just a few years ago when "Scientists across the world have declared that no further research is required in the field of astronomy, saying that everything about space is now sufficiently understood" \citep{Cosmos2021}.
 
 In Section~\ref{Methods} we summarize our approach to determine if creativity has declined over time by using constellation complexity as our metric. In Section~\ref{Data} we provide an overview of our constellation data, and explain how we determine the complexity of these constellations in Section~\ref{Complexity}. We demonstrate how this complexity has decreased over time in Section~\ref{Trends}. We then summarize this paper in Section~\ref{Summary}.

\section{Methods} \label{Methods}

\subsection{Data} \label{Data}
To begin our study of constellation complexity, we must first define which constellations we are discussing. In particular, we focus on the 88 constellations established by the IAU \citep{Delporte1930} and also frequently utilize three-letter abbreviations of those constellations, for much the same reason as Russell initially criticised using full constellation names \citep{Russell1922}:
\begin{quote}
    the long Latin names of the constellations have been printed in full (or somewhat curtailed) thousands upon thousands of times, in astronomical publications, when similar abbreviations would have answered every purpose. The aggregate cost of the extra printing involved must have been very considerable, and there is no reason for the continuance of this needless expense, especially in the present day of high prices.
\end{quote}

For all 88 we then encode ages for all constellations, which fall into a handful of ages as outlined in Table~\ref{tab1:Ages}. We are beginning with a starting source of ages as listed on Wikipedia \footnote{\url{https://en.wikipedia.org/wiki/IAU_designated_constellations}}, but have attempted to further refine those ages given the known issues with Wikipedia accuracy, particularly when it comes to historical articles \citep{Holman2008}.

\begin{table*}[ht]
\caption{Ages of Constellations}
\label{tab1:Ages}
\begin{tabular}{p{0.1\linewidth}| p{0.5\linewidth}|p{0.15\linewidth}|p{0.15\linewidth}}  \hline
Year & Constellations & Wiki Source & Citation \\ \hline
ancient & And, Aql, Aqr, Ara, Ari, Aur, Boo, Cap, Cas, Cen, Cep, Cet, CMa, CMi, Cnc, CrA, CrB, Crt, Crv, Cyg, Del, Dra, Eqi, Eri, Gem, Her, Hya, Leo, Lib, Lup, Lyr, Oph, Ori, Peg, Per, PsA, Psc, Sco, Ser, Sge, Sgr, Tau, Tri, UMa, UMi, Vir & Ptolemy & \citep{Ptolemy} \\ \hline
1536 & Com & Caspar Vopel & \citep{Vopel1536, Dekker2010}  \\ \hline
1589-1613 & Cru, Col, Cam, Mon & Plancius & \citep{Plancius1592, Plancius1613} \\ \hline
1603 & Aps, Cha, Dor, Gru, Hyi, Ind, Mus, Pav, Phe, TrA, Tuc, Vol & Uranometria & \citep{Bayer1603} \\ \hline
1690 & CVn, Lac, LMi, Lyn, Sct, Sex, Vul & Hevelius & \citep{Hevelius1690} \\ \hline
1763 & Ant, Cae, Cir, For, Hor, Mic, Nor, Oct, Pyx, Ret, Scl, Tel, Pic, Men, Car, Pup, Vel & Lacaille & \citep{delaCaille1763} \\ \hline
\end{tabular}
\end{table*}

For the 48 Ptolemaic constellations we do not know precise ages but their appearance in the star catalogue included in the \textit{Algamest} by \citet{Ptolemy} gives them a definite constraint of having been established by around the year 150 --- though they are all likely much older, and some contentious claims put some of these constellations at dating back tens of thousands of years \citep{BBC2000, Sparavigna2008, Faris2018}. In any case, we simply consider these to be our initial set of 'ancient' constellations.

The first constellation with any sense of a refined discovery time does not come until the 16th century, with the addition of Coma Berenices. Sourced to the cartographer and instrument maker Caspar Vopel, the identification of Com as a constellation comes not from an atlas or catalogue, but from a celestial globe he had made \citep{Vopel1536, Dekker2010}.

The next group of sixteen constellations is somewhat muddied in terms of discovery, but Wikipedia attributes them to Petrus Plancius and to Johannes Bayer's \textit{Uranometria}, with a range of dates between 1589 and 1613. The bulk of these (Aps, Cha, Dor, Gru, Hyi, Ind, Mus, Pav, Phe, TrA,
Tuc, Vol) are attributed to \citet{Bayer1603} and his \textit{Uranometria}, however it appears that this is just their first appearance in a celestial atlas. There is some strong indication that, prior to 1603, these constellations were actually identified, following the instructions of Plancius, by Pieter Dirkszoon Keyser (sometimes referred to as Petrus Theodorus, and in Knobel as Pieter Dircksz Keyzer) in 1596 and 1597 \citep{Knobel1917}. Knobel goes so far as to summarize his study of records relating to observations of these twelve constellations as follows:
\begin{quote}
    Though all the evidence is circumstantial, the following inferences are irresistible: that the whole catalogue and the formation of the new twelve constellations must be attributed to Pieter Dircksz Keyzer, and not in any way to Frederick de Houtman: that the catalogue was sent back by the \textit{Hollandia}, and after the return of that ship on 1597 August 10 it would go to Plancius, who probably communicated it to Bayer: and that Frederick de Houtman obtained an imperfect copy of it which, as Pieter Dircksz was dead, he published as his own work
\end{quote}
These twelve constellations would then be published on a celestial globe by Plancius in approximately 1598 \citep{Plancius1598}. This also means that they join four other constellations as seeming to have been first published in works by Petrus Plancius, with map being the first published appearances for Cru and Col \citep{Plancius1592}, and another celestial globe as the source for Cam and Mon \citep{Plancius1613}.
There is a case to be made that these constellations can be pushed to slightly earlier dates for Western observers, such as observations of the modern constellation of Crux in 1500 by Jo{\~a}o Faras, then calling it "las guardas", but we are constraining ourselves to published appearances for the scope of this work \citep{Varnhagen1843}.

The next batch of constellations comes from the posthumous publication of  Johannes Hevelius' \textit{Firmamentum Sobiescianum, sive Uranographia}, typically considered associated with his larger \textit{Prodromus Astronomiae}, which featured a star catalogue based on his observations --- the publication of all this work was arranged by his wife and assistant Elisabeth Hevelius \citep{Hevelius1690}.

The final group of constellations to be published and subsequently included in the modern IAU list of constellations is a group attributed to Nicolas-Louis de Lacaille (or de La Caille), most of which were poshumously published in his catalogue of stars after observing the Southern sky from 1750 to 1754 from the Cape of Good Hope \citep{delaCaille1763}.
There are three constellations in this group that were published slightly earlier, as they had originally been part of the Ptolemaic constellation of Argo or Argo Nevis, but Lacaille suggested dividing the large constellation into smaller the smaller regions of Car, Pul, and Vel as Argo contained over 160 easily distinguishable stars\citep{LaCaille1756}. This issue would not be settled until the IAU definitions, but would see gradually increasing support over the intervening period \citep{Hershel1844}.

Collectively these 88 constellations, as defined by the IAU and with their constituent stars, are available in a JSON file that is provided within the Stellarium open-source planetarium software \citep{Stellarium}. Specifically to our usage, each constellation has included with it a list of lists, where each list is a set of stars (using HIPPARCOS identifiers) that can be used to draw a continuous line to form the constellation. It is this data that we use when we move on to address constellation complexity in the following subsection.

\subsection{Complexity} \label{Complexity}
As constellations are fundamentally defined as patterns made up from connecting bright stars in the night sky (in a Western context, at least - see work such as \citep{Gullberg2020} for a discussion of how other cultures feature 'dark constellations'). This makes constellations an analogue of graphs, and so we can leverage graph theory to determine the cyclomatic number of constellations to assess the complexity of constellations \citep{Berge2001, Mccabe1976}. We can use the following equation to calculate complexity, M, where N is the number of stars, E is the number of connecting segments between stars, and P is the number of connected components (where a constellation where all stars are connected to a single network has a a single connected component):
\begin{equation}
\label{eq:complexity}
M = E - N + 2P
\end{equation}

\begin{figure*}[!htb]
    \begin{subfigure}{0.45\textwidth}
    \includegraphics[width=1\textwidth]{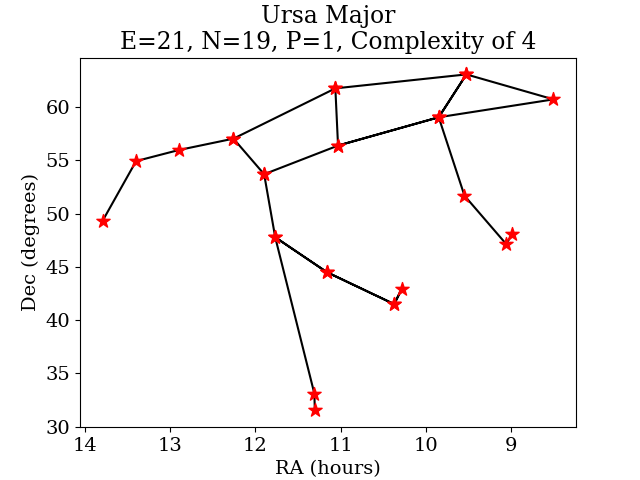}
    \caption{Ursa Major = ancient constellation}
    \end{subfigure}
    \hfill
    \begin{subfigure}{0.45\textwidth}
    \includegraphics[width=1\textwidth]{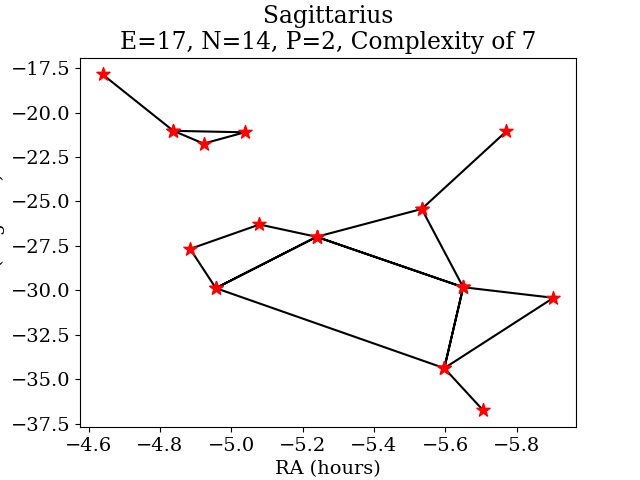}
    \caption{Sagittarius - ancient constellation}
    \end{subfigure}
    \hfill

    \medskip
    \begin{subfigure}{0.45\textwidth}
    \includegraphics[width=1\textwidth]{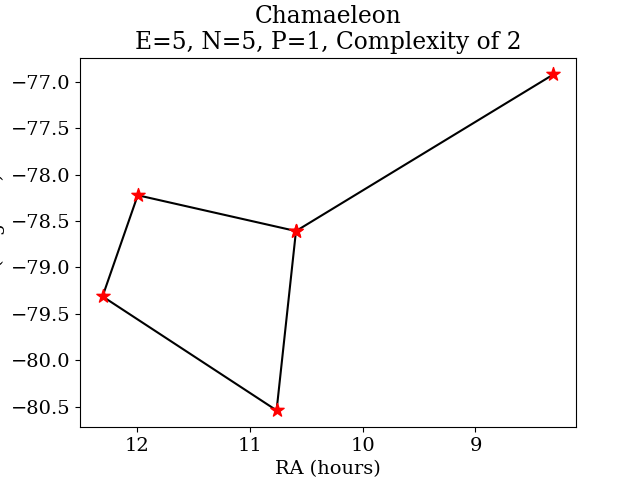}
    \caption{Chamaeleon - circa 1600 constellation}
    \end{subfigure} 
    \hfill 
    \begin{subfigure}{0.45\textwidth}
    \includegraphics[width=1\textwidth]{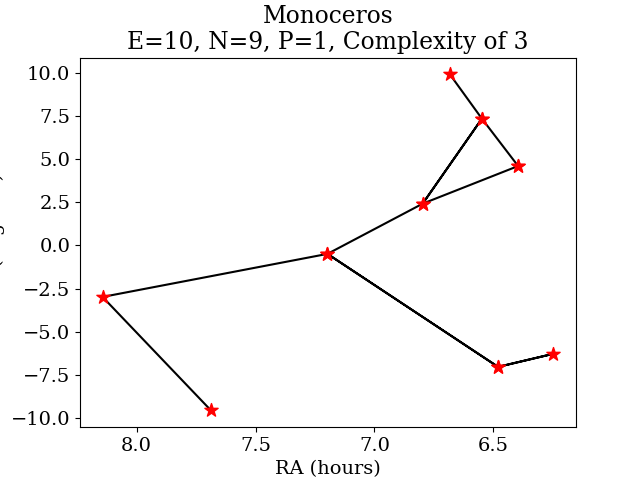}
    \caption{Monoceros - circa 1600 constellation}
    \end{subfigure}
    \hfill

    \medskip
    \begin{subfigure}{0.45\textwidth}
    \includegraphics[width=1\textwidth]{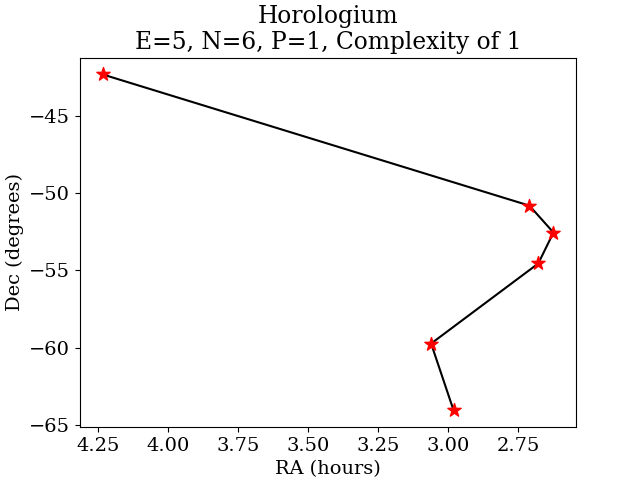}
    \caption{Horologium - 1763 constellation}
    \end{subfigure}
    \hfill  
    \begin{subfigure}{0.45\textwidth}
    \includegraphics[width=1\textwidth]{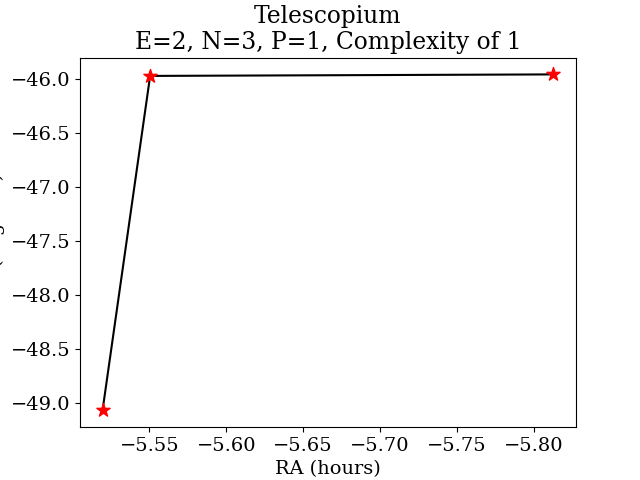}
    \caption{Telescopium - 1763 constellation}
    \end{subfigure}

  \caption{Six constellations with their complexity metrics included. This represents two ancient constellations, two constellations from around 1600, and two constellations from the 1700s. All nodes must be the locations of stars in the constellation, shown here as red stars in each figure.}
  \label{fig:constellations}
\end{figure*}

Using the constellation data that we have sourced from Stellarium, we then calculated the complexity for all 88 IAU constellations. We provide a few examples of the charted constellation shapes in Figure~\ref{fig:constellations} by combining the information in Stellarium with the coordinates of the stars based on their HIPPARCOS numbers \citep{Perryman1997}. With this metric outlined in Equation~\ref{eq:complexity}, the simplest constellation would have a complexity metric of 1.

\subsection{Trends over Time} \label{Trends}
In order to look at trends over time, we divide the IAU constellations into three groups based on when they were established, roughly folowing the years listed in Table~\ref{tab1:Ages}. The first group is the 47 ancient constellations, which date back at least as far as Ptolemy (and likely much older). The second group is the 24 constellations that were dated to the 1500s and 1600s. The third group is the 17 constellations that were dated after 1700. There are, as noted in Section~\ref{Data}, some variations between the ages of these constellations in Table~\ref{tab1:Ages} and the true ages, but we have chosen breaks in time that are sufficiently far spaced out so as not to be impacted.

\begin{figure*}[!htb]
    \begin{center}
   \includegraphics[width=0.95\textwidth]{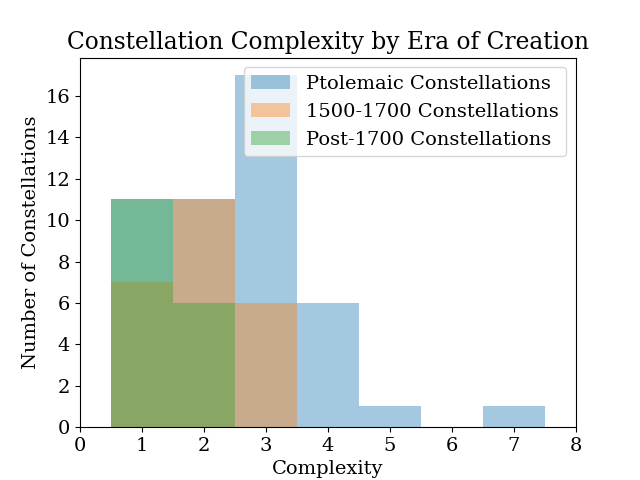}
  \end{center}
\caption{Histogram of the complexity metric for all constellations when broken into three groups: ancient (Ptolemaic) constellations, constellations from the 1500s and 1600s, and constellations from after 1700.}
\label{fig:trends}
\end{figure*}

When we compare these three groups, we see that both the median and mean constellation complexity decrease each time we move to a younger group of constellations. The ancient constellations have a median complexity of 3 and a mean complexity of 2.6, the 1500s and 1600s constellations have a median complexity of 2 and a mean complexity of 2.0, and the post-1700 constellations have a median complexity of 1 and a mean complexity of 1.4. We also plot these three groups as histograms in in Figure~\ref{fig:trends}, which visually shows that the distribution of constellation complexity shifts to lower values as we move to younger constellations.

It would be a reasonable question to ask if this simply is an artifact of small-number statistics as our groups of constellations do overall and we are dealing with only 88 constellations spread across three groups. In order to compare these groups, we use a Mann-Whitney U rank test to test the null hypothesis where the complexity of two groups of constellations have the same distribution \citep{Mann1947}.

When we assess our results from the Mann-Whitney U rank test (and using a permutation test with 100,000 samples), we find that just comparing the Ptolemaic constellations to all post-1500 constellations we get a p-value of 0.00042, lettings us reject the same underlying distribution with a greater than 99\% confidence. With slightly more granularity, we also find a p-value of 0.043 when comparing the Ptolemaic constellations and the 1500-1700s constellations, allowing us to reject the null hypothesis for those two groups with 95\% confidence. Finally, when we compare the 1500-1700s constellations and the post-1700s constellations, the p-value is 0.11, again allowing us to reject the null hypothesis again with a greater than 95\% confidence. In every case, we find that the complexity of older constellations is statistically significantly higher than the complexity of younger constellations.

\section{Summary}\label{Summary}
In this paper we have determined the complexity of all 88 modern IAU constellations and used this as a proxy for the creativity of the astronomers that defined those constitutions, showing that older constellations feature greater complexity and therefore reflect a more creative society. It is of no surprise that constellations that have had rich mythologies built around them reflect greater creativity than just looking at things in the office and saying that they're constellations \citep{Hamilton1942, McKay2004}. There may be ways to make astronomy a more creative field, however we consider that to be beyond the scope of this paper as we cannot think of what those ways would be.

\section{Acknowledgments}
This work made use of Astropy:\footnote{http://www.astropy.org} a community-developed core Python package and an ecosystem of tools and resources for astronomy \citep{astropy2013, astropy2018, astropy2022}.

This research has made use of NASA’s Astrophysics Data System.

Software: astropy \citep{astropy2013, astropy2018, astropy2022}, matplotlib \citep{Hunter2007}, numpy \citep{Harris2020}, pandas \citep{Pandas2020, Mckinney2010}, scipy \citep{Virtanen2020}, Stellarium \citep{Stellarium}


\bibliographystyle{apalike}
\bibliography{main}


\end{multicols}

\end{document}